\documentclass{article}

\usepackage[numbers]{natbib}         
\usepackage[colorlinks]{hyperref}    
\usepackage[english]{babel} 
\usepackage{amssymb}
\usepackage{amsmath}
\usepackage{txfonts}
\usepackage{mathdots}
\usepackage[classicReIm]{kpfonts}
\usepackage[dvips]{graphicx} 
\usepackage[a4paper, portrait, margin=1in]{geometry}
\usepackage{tabularx}
\usepackage{longtable}
\usepackage{multirow}
\usepackage{booktabs}
\usepackage[labelsep=period]{caption}
\usepackage{makecell}
\begin{document}
	\setlength{\parindent}{0pt}
	\setlength{\parskip}{1ex}
	
	\textbf{\Large Deep learning-based Real-time Volumetric Imaging for Lung Stereotactic Body Radiation Therapy: A Proof of Concept Study}
	
	\bigbreak

	Yang Lei$^+$, Zhen Tian$^+$, Tonghe Wang, Kristin Higgins, Jeffrey D. Bradley, Walter J. Curran, Tian Liu and Xiaofeng Yang*
	
	Department of Radiation Oncology and Winship Cancer Institute, Emory University, Atlanta, GA 30322
	
	$^+$Co-first author

	\bigbreak
	\bigbreak
	\bigbreak

	\textbf{*Corresponding author: }
	
	Xiaofeng Yang, PhD
	
	Department of Radiation Oncology
	
	Emory University School of Medicine
	
	1365 Clifton Road NE
	
	Atlanta, GA 30322
	
	E-mail: xiaofeng.yang@emory.edu

	\bigbreak
	\bigbreak
	\bigbreak
	\bigbreak
	\bigbreak
	\bigbreak

	\textbf{Abstract}

	Due to the inter- and intra- variation of respiratory motion, it is highly desired to provide real-time volumetric images during the treatment delivery of lung stereotactic body radiation therapy (SBRT) for accurate and active motion management. In this proof-of-concept study, we propose a novel generative adversarial network integrated with perceptual supervision to derive instantaneous volumetric images from a single 2D projection. Our proposed network, named TransNet, consists of three modules, i.e., encoding, transformation and decoding modules. Rather than only using image distance loss between the generated 3D images and the ground truth 3D CT images to supervise the network, perceptual loss in feature space is integrated into loss function to force the TransNet to yield accurate lung boundary. Adversarial supervision is also used to improve the realism of generated 3D images. We conducted a simulation study on 20 patient cases, who had received lung SBRT treatments in our institution and undergone 4D-CT simulation, and evaluated the efficacy and consistency of our method for four different projection angles, i.e., 0°, 30°, 60° and 90°. For each 3D CT image set of a breathing phase, we simulated its 2D projections at these angles. Then for each projection angle, a patient’s 3D CT images of 9 phases and the corresponding 2D projection data were used for training, with the remaining phase used for testing. The mean absolute error (MAE), normalized MAE (NMAE), peak signal-to-noise ratio (PSNR) and structural similarity index metric (SSIM) achieved by our method are 99.3±14.1 HU, 0.032±0.007, 23.4±2.2 dB and 0.949±0.012, respectively. These results demonstrate the feasibility and efficacy of our 2D-to-3D method for lung cancer patients, which provides a potential solution for in-treatment real-time on-board volumetric imaging for accurate dose delivery to ensure the effectiveness of lung SBRT treatment.
	
	\bigbreak
	\bigbreak
	
	\textbf{keywords:} Lung stereotactic body radiation therapy, volumetric imaging, deep learning, conditional generative adversarial network, perceptual supervision.

	\noindent 
	\section{ INTRODUCTION}
	
	Respiratory motion causes substantial anatomic changes, leading to significant dosimetric uncertainties in lung cancer radiotherapy \cite{RN3, RN4}. The impact of these uncertainties is amplified in lung stereotactic body radiation therapy (SBRT), which delivers very high and conformal doses in relatively few fractions \cite{RN5, RN7}. Motion management is hence essential for treatment accuracy and planning margin reduction in lung SBRT \cite{RN6, RN3}. Over the years, time-resolved imaging modalities have been developed and adopted clinically for motion management, such as 4DCT which is routinely used for treatment planning \cite{RN4, RN16967, RN3}, and 4D cone beam CT which is an emerging imaging tool to aid patient positioning \cite{RN2244, RN2324, RN2325}. Nevertheless, respiratory motion can vary between CT simulation, patient positioning and treatment delivery in its magnitude, baseline, period and regularity \cite{RN16968}. Hence, in-treatment real-time imaging is highly desired to provide the actual anatomy information during treatment delivery for in-treatment motion monitoring and management \cite{RN4, RN8}. 
	
	Projection based 2D imaging, such as kV fluoroscopy and MV Beam Eye View images is a common way to realize in-treatment real-time imaging \cite{RN688}. However, because the patient’s 3D anatomy is projected and superimposed onto a 2D plane, the visibility of the tumor is very limited, especially when it is behind a high intensity structure such as bone \cite{RN8}. The on-board imaging component of MRIdian MRI-guided linear accelerator (ViewRay company, Cleveland, Ohio) is a pioneering system that can perform real-time imaging in a single 2D pane without the aforementioned superimposing issue \cite{RN18}. However, the unobserved tumor/OAR motion in other planes during treatment delivery might impair the tracking accuracy \cite{RN19}, let alone the high cost of this system and the interference between its magnetic field and the radiation beams. Another real-time 2D imaging scheme has also been proposed to capture the instantaneous anatomy on a 2D plane by measuring the scatters of the therapeutic beams during treatment, which demands a super sensitive detector to capture the relatively small number of scattered photons and has the same issue of unobserved motion on other plans as cine-MRI \cite{RN8}.
	
	Because of the wide availability of on-board imaging system integrated in Linac machines, it is highly desired to generate instantaneous volumetric image using this on-board imager. Due to the slow rotation of Linac gantry, it is impossible to acquire sufficient angular projection data at high frequency to satisfy the Shannon-Nyquist sampling theorem for traditional computed tomography or tomosynthesis approaches to derive the instantaneous volumetric image \cite{RN1}. Although compressed sensing techniques have been employed to successfully reduce the required number of projections to achieve low dose CBCT \cite{RN20, RN21}, the resultant sparsity is far away from meeting the demand of real-time volumetric imaging \cite{RN1}. Recently, efforts have been made to derive a volumetric image from a single x-ray projection either using 2D-3D registration techniques \cite{RN562, RN563} or machine learning techniques \cite{RN24, RN25}, in combination with a lung motion model as a prior knowledge. However, the robustness of these techniques to handle the variation in respiratory motion, patient positioning and anatomy change is still a concern for clinical application \cite{RN8}. 
	
	Deep learning techniques have attracted much attention for their ability to learn complex relationships and to incorporate existing knowledge into the inference model \cite{RN26, RN27}, and have found widespread applications across different disciplines \cite{RN29, RN31, RN30, RN32, RN28}. Shen et al. has recently trained a deep neural network to learn the feature-space transformation between a 2D projection and a 3D volumetric CT image, and used this trained model to derive volumetric CT image from an anterior-posterior single projection view \cite{RN1}. Inspired by this novel idea, we propose to employ the advanced perceptually supervised Generative Adversarial Networks (GAN)-based approach to generate volumetric image from a single projection and have tested its effectiveness and robustness for different projection angles. This approach has 2 distinctive strengths: 1) By adding adversarial loss to differentiate the distribution of generated 3D volume from ground truth 3D volume, the capability of our method to generate more realistic 3D CT is significantly improved. 2) By adding perceptual loss in feature space, our networks are enforced to yield more accurate contrast around lung organ. This proof-of-concept study is expected to pave the way for real-time volumetric imaging for more accurate and active motion management during treatment delivery of lung SBRT treatment.

	\noindent 
	\section{Methods and Materials}
	\noindent 
	\subsection{System overview and network design}
	
	Our implementation of the transformation from 2D projection to 3D volumetric image consists of two major stages: a training stage and a transform stage. Fig. 1 shows a schematic flow chart of the training stage. At this stage, a feed forward path, named TransNet, is trained under several supervision mechanisms that perform learnable parameters optimization to learn the 2D-to-3D mapping. A set of paired 2D projection (with angle n° and size of $h_1\times w_1$) and 3D CT volume (with size of $h_2\times w_2\times d$) is served as the training data set. TransNet takes a single 2D projection as input, and outputs a generated synthetic 3D CT. The corresponding real 3D CT is used as ground truth to supervise TransNet. 
	
	TransNet consists of three subnetworks, i.e., an encoding module, a transformation module and a decoding module. The encoding module is to explore 2D features from the 2D projection data. It is composed of several 2D residual blocks \cite{RN2} and three 2D max pooling layers to reduce the feature map size to $\frac{h_1}{8}\times\frac{w_1}{8}\times{8\bullet n}_f$. Here, $n_f$ denotes the number of feature maps extracted from the first residual block of encoding module. After this module, the number of feature maps is increased to ${8\bullet n}_f$. The following transformation module is to bridge the feature maps from 2D scene to 3D scene. It uses tensor reshaping to transform the 2D feature map to the corresponding 3D one), and employs several 3D convolution layers with valid padding and a tensor bicubic interpolation to resize the transformed 3D feature map to a size of $\frac{h_2}{8}\times\frac{w_2}{8}\times\frac{d}{8}\times int\left(\frac{{8\bullet n}_f}{d}\right))$. The decoding module then generates transformed 3D CT image (with a size of $h_2\times w_2\times d$) from these 3D feature maps. This module is composed of several 3D deconvolution layers, each followed with a 3D convolution layer and an extra 3D convolution layer to perform regression. 
	
	The transform stage follows the same feed forward path in the training stage, that is, extracting 2D feature from the input 2D projection and feeding it into TransNet to derive the 3D CT volume.
	
	\begin{figure}
		\begin{center}
		\noindent \includegraphics*[width=6.50in, height=4.20in, keepaspectratio=true]{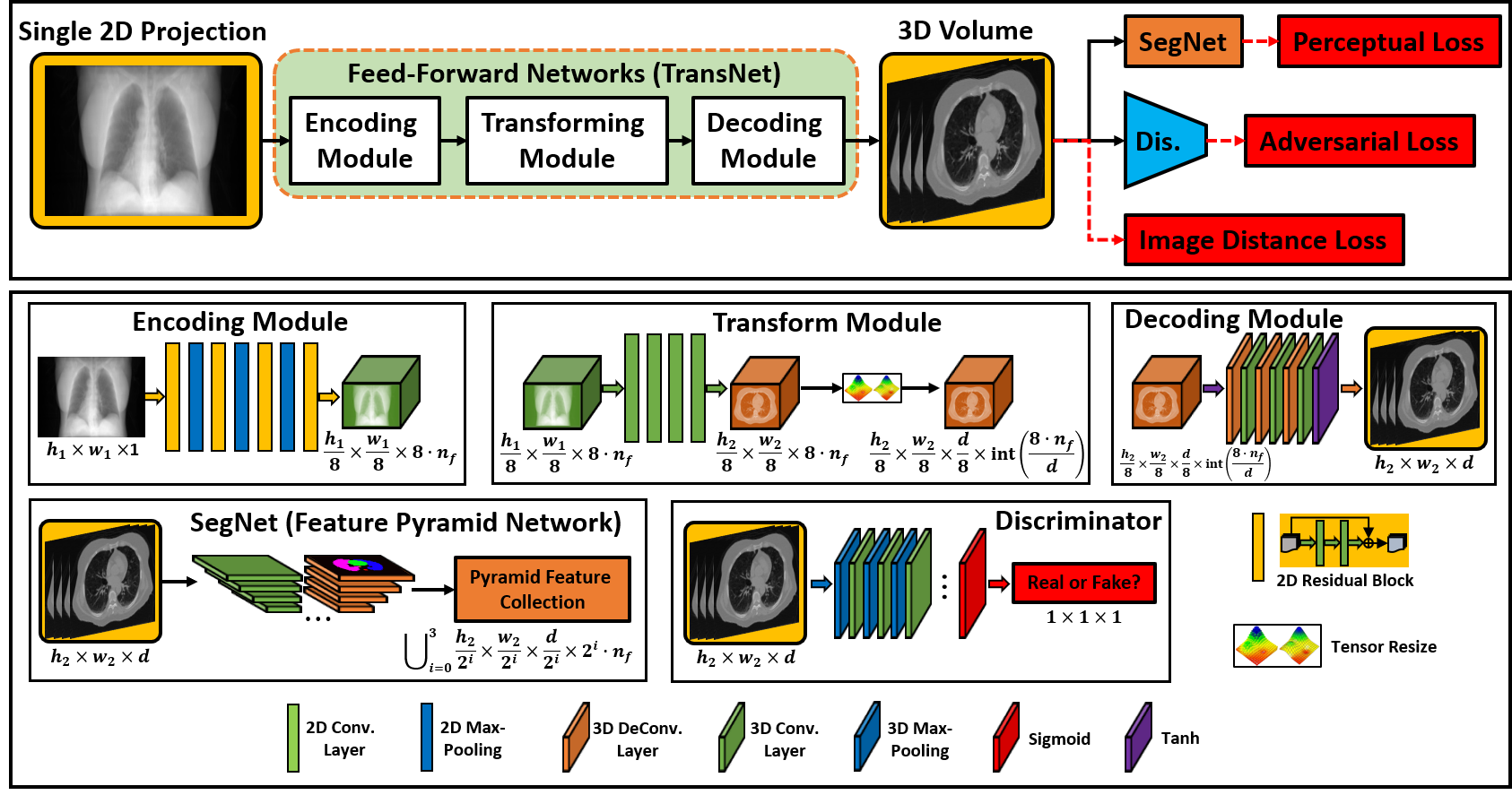}
		\end{center}
		
		\noindent Fig. 1. Schematic flow chart of the proposed TransNet for 2D-to-3D generation. The top box shows the framework of TransNet. The bottom one shows the architecture of each module used in TransNet. SegNet and discriminator shown in this figure will be present in subsection 2.3 and 2.4, respectively.
	\end{figure}

	\noindent 
	\subsection{Supervision and loss function}
	
	The performance of this proposed TransNet heavily relies on the loss function used for supervision, which is formulated as:
	\begin{equation} 
		L_{total}\left(\hat{y},y\right)={\lambda_{mae}\bullet L}_{mae}\left(\hat{y},y\right)+{\lambda_{gd}\bullet L}_{gd}\left(\hat{y},y\right){\ +{{\ \lambda}_p\bullet L}_p\left({f_{\hat{y}},f}_y\right)+L}_{adv}\left(\hat{y}\right)
	\end{equation} 
	in our study. Here, $L_{mae}\left(\hat{y},y\right)$ denotes the mean absolute error (MAE) between transformed 3D CT $\hat{y}$ and ground truth 3D CT y to measure the voxel-level error. $L_{gd}\left(\hat{y},y\right)$ denotes the gradient difference (GD) to measure the gradient-level error. These two terms are used as image distance loss between the transformed 3D CT and the corresponding ground truth 3D CT. $L_p\left({f_{\hat{y}},f}_y\right)$ denotes the perceptual loss, which is defined in feature domain instead of image domain and calculates the difference between the feature map of the transformed 3D CT (i.e., $f_{\hat{y}}$) and the feature map of the ground truth (i.e., $f_y$). The last term, ${\ L}_{adv}\left(\hat{y}\right)$, denotes the adversarial loss, which is deployed to force TransNet to yield realistic 3D images. This perceptual loss and adversarial loss will be present in details in subsection 2.3 and 2.4, respectively. $\lambda_p$, $\lambda_{mae}$ and $\lambda_{gd}$ denote the balancing hyper-parameters of perceptual, MAE and GD loss, respectively. TransNet is trained via minimizing this total loss.

	\noindent 
	\subsection{Perceptual loss}
	
	Mapping a single 2D projection to the much more informative 3D image is a very ill-posed problem. If we only use the image distance loss (e.g., MAE and GD), it is very challenging for TransNet to well represent the correlation between the image difference in 2D projection induced by respiratory motion and the corresponding image difference in 3D volume, resulting in inaccurate or blurry  boundary of the moving organs. Since lung is one of the organs that are most susceptible to respiratory motion in thoracic CT images, and lung toxicity is one of the major concerns for lung SBRT treatment, we therefore proposed to integrate perceptual supervision into our networks in our study to enforce more power on generating accurate lung boundary in the transformed 3D images. 
	
	The main idea of perceptual supervision \cite{RN3} is that feed forward networks (i.e., TransNet in this work) could generate high-confidence fooling image (i.e., transformed 3D CT in this work) by using a perceptual loss that measures the perceptual and semantic difference between transformed 3D volume and ground truth 3D volume. In our study, we define the perceptual loss as the feature difference on high-level feature maps. These high-level feature maps are extracted from both transformed 3D volume and ground truth 3D volume, via a network named SegNet pretrained for image semantic segmentation. As shown in Fig.1, SegNet uses the traditional end-to-end feature pyramid network (FPN) architecture, and extracts pyramid feature maps from both transformed and ground truth 3D volumes for perceptual supervision. It was pre-trained using the dataset of chest CT images and paired lung contours obtained from 2017 AAPM Thoracic Auto-segmentation Challenge \cite{RN5, RN6}. 
	  
	Denoting TransNet as $F_e\circ F_t\circ F_d$, the transformation of an input 2D projection x to a 3D volume $\hat{y}$  can be denoted by $\hat{y}=F_d\left(F_t\left(F_e\left(x\right)\right)\right)$. SegNet, denoted by $F_s$, extracts multi-level feature maps from the 3D ground truth CT (y) and the transformed one ($\hat{y}$), respectively, i.e., $f_y=\bigcup_{i=1}^{N}{F_s^i\left(y\right)}$ and $f_{\hat{y}}=\bigcup_{i=1}^{N}{F_s^i\left(\hat{y}\right)}$, where $N$ is the number of pyramid levels. The perceptual loss is defined as the Euclidean distance between the two feature maps, calculated as 
	\begin{equation} 
	L_p\left(f_y,f_{\hat{y}}\right)=\sum_{i=1}^{N}{\frac{\omega_{i}}{C_{i}\bullet H_{i} \bullet W_{i} \bullet D_{i}}}\left|\left|F_{s}^{i}(y)-F_{s}^{i}(\hat{y})\right|\right|_{2}^{2}                    
	\end{equation} 
	where $C_i$ denotes the number of feature map channels at ith pyramid level. $H_i$, $W_i$ and $D_i$ denotes the height, width and depth of that feature map. According to Fig. 1, these values are calculated as $C_i={2^i\bullet n}_f$, $H_i=\frac{h_2}{2^i}$, $W_i=\frac{w_2}{2^i}$, and $D_i=\frac{d}{2^i}$. $\omega_i$ is a balancing parameter for feature level $i$. Since the semantic information of the feature map at higher pyramid levels would be coarse, the weight for that level’s perceptual loss should be enlarged, thus we chose $\omega_i={1.2}^{i-1}$.

	\noindent 
	\subsection{Adversarial loss}
	
	Apart from accurate lung boundary, another requirement on the derived synthetic 3D images is to be realistic compared to the ground truth images. Due to the motion artifact, the ill-posed TransNet would introduce additional noise or bias into the transformed 3D CT. Our aim is to reduce this artificial noise via enhancing the realism of transformed 3D CT. 
	
	In previous conditional generative adversarial networks (CGAN) studies \cite{RN2, RN7}, the common way to improve the output’s realism is to use a discriminator network to discriminate the synthetic one (i.e., transformed 3D CT in this task) from real one (i.e., ground truth 3D CT in this task). The discriminator used in our study is a traditional fully convolution network (FCN). It is composed of several 3D convolution layers and 3D max-pooling layers to reduce the size of feature map to 1×1×1×1, and a sigmoid layer to binarize the output (i.e., 1 means that the discriminator think the input is real, and 0 means synthetic). Denoting the discriminator as $F_{dis}$, the discriminator loss is to measure its recognition error, calculated as
	
	\begin{equation} 
	L_{dis}\left(\hat{y},y\right)=SCE\left(F_{dis}\left(\hat{y}\right),0\right)+SCE\left(F_{dis}\left(y\right),1\right)                      
	\end{equation}
	
	Here, $F_{dis}\left(\bullet\right)$ is a binary value indicating whether the discriminator determines the input to be real or synthetic image. The function $SCE\left(\bullet,\bullet\right)$ is the sigmoid cross entropy between two distributions. The more similar the two distributions are, the smaller the value of sigmoid cross entropy is. The adversarial loss is to measure the recognition accuracy for the 3D images generated by TransNet, calculated as
	
	\begin{equation}
		L_{adv}\left(\hat{y}\right)=SCE\left(F_{dis}\left(\hat{y}\right),1\right)                
	\end{equation}
	By minimizing both discriminator’s loss (i.e., minimizing the recognition error of differentiating synthetic images from real images) and generator’s adversarial loss (i.e., maximizing the recognition error for high-quality 3D image that can fool the discriminator), TransNet can yield a 3D volume of a comparable realistic level with the ground truth 3D volume. 
	
	\noindent 
	\subsection{Training and testing}
	
	Our method was implemented in python 3.6.8 and TensorFlow 1.8.0 on a NVIDIA Tesla V100 GPU with 32GB of memory. Adam gradient optimizer with learning rate of 2e-4 was used for optimization. Currently, we do not have in-treatment instantaneous volumetric image technically available for us to use as ground truth. Hence, in this proof-of-concept study, we use 4D CT images of our previous lung cancer patients, which were acquired for treatment planning, to test the feasibility of the proposed perceptually supervised GAN-based volumetric imaging method. 4D CT images of 20 previous patients were collected and anonymized. Institutional review broad approval was obtained; informed consent was not required for this Health Insurance Portability and Accountability Act (HIPPA)-complaint retrospective analysis. All these 4D CT images were acquired on a Siemens SOMATOM Definition AS CT scanner with a resolution of 0.977×0.977×2 mm$^{3}$. Each 4D CT image consists of 10 3D CT sets, corresponding to 10 respiratory phases. 
	
	In our study, apart from the anterior-posterior projection angle (i.e., 0°), we have also tested three other projection angles 30°, 60° or 90°, respectively. For each 3D CT image set, 2D projection data at these angles were generated by forward-projecting the CT volume to a plane using the same geometry as an on-board imager of Varian (Palo Alto, CA) Truebeam Linac. 40 experiments were performed for each patient, varying the combination of phase and projection angle. Specifically, at each experiment, our networks were trained and tested for one of the four projection angles (0°, 30°, 60° or 90°) in one phase. The pairs of 2D projection and the corresponding 3D CT image of 9 breathing phases were used as the training data, and the remaining phase was used for testing. We repeated this experiment 40 times to let each angle with each phase used as test data exactly once. To further enlarge the variance of training data, data augmentation, which includes rotation, flipping and scaling, was applied. 
	
	Four metrics were used in our study for quantitative comparison, that is, mean absolute error (MAE), normalized MAE (NMAE), peak signal-to-noise ratio (PSNR) and structural similarity index metric (SSIM). MAE is a magnitude of the absolute difference between the transformed 3D CT and ground truth 3D CT. NMAE is a magnitude of the absolute difference between the normalized transformed 3D CT and normalized ground truth 3D CT (normalized images to [0, 1] using a same normalization factor). PSNR is used to compare the noise level between the transformed 3D CT and ground truth 3D CT. SSIM, which is commonly used in pattern matching and image analysis, is a measure of the similarity of image structures. The calculations of these metrics were conducted over the entire image volume.
	
	To demonstrate the contribution of perceptual supervision and adversarial supervision to the final performance, a step-by-step study was performed to evaluate the effectiveness of each component. In order to illustrate the statistical significance of quantitative improvement by these steps, paired two-tailed t-tests were also performed among 10 phases for each patient.

	\noindent 
	\section{RESULTS}
	\noindent 
	\subsection{Efficacy of perceptual supervision}
	
	To demonstrate the efficacy of the perceptual supervision, we compared the transformation accuracy of the proposed TransNet between two different supervision strategies, i.e., 1) being supervised by MAE and GD losses (referred to as TransNet-v1), and 2) being supervised by a combination of MAE, GD and perceptual losses (referred to as TransNet-v2). This comparison was conducted on all the patients at one projection angle (60°) for demonstration purpose.
	
	Fig. 2 shows the visual comparison between TransNet-v1 and TransNet-v2 for a patient. Axial views of the transformed 3D CTs of these two methods are shown in subfigure (a2) and (a3), respectively, with the corresponding zoomed-in images in (b2) and (b3). The ground truth image and zoomed-in image are shown in subfigure (a1) and (b1). The corresponding feature maps that were extracted via SegNet are shown in subfigures (c1)-(c3), with the corresponding zoomed-in feature map shown in (d1)-(d3). It can be observed from these images that the feature map of TransNet-v2 around lung boundary is more similar to that of ground truth, which leads to less blurry and more accurate lung boundary in TransNet-v2 than TransNet-v1. From the line profile of yellow line on ground truth CT, TransNet-v1 and TransNet-v2 generated 3D CTs (shown in e2), we can see that the generated 3D volume via TransNet-v2 has a more accurate lung contrast as compared to TransNet-v1.

	\begin{figure}
		\begin{center}
		\noindent \includegraphics*[width=6.50in, height=4.20in, keepaspectratio=true]{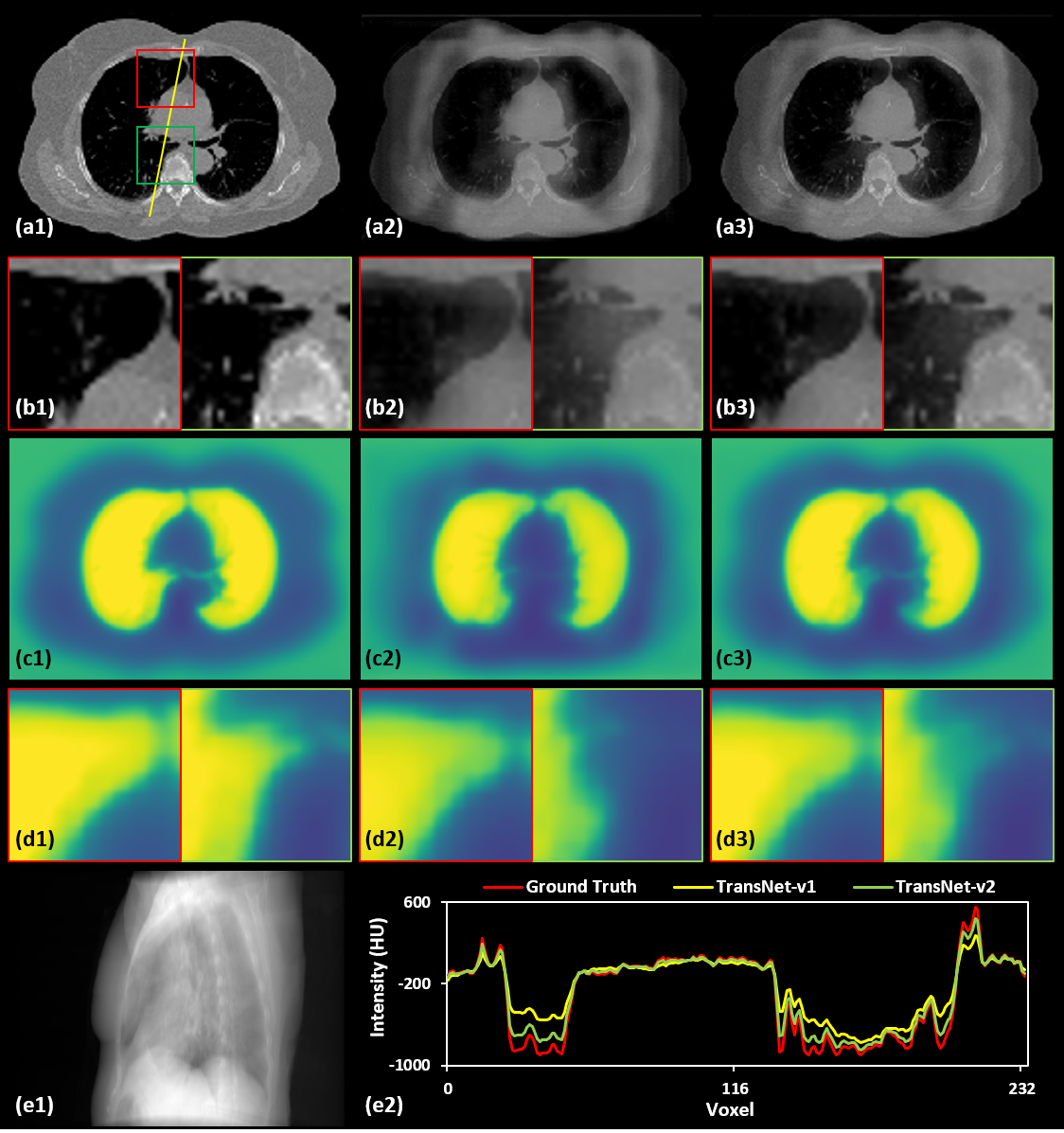}
		\end{center}
		
		\noindent Fig.2 Visual comparison between TransNet-v1 and TransNet-v2 to demonstrate the efficacy of perceptual supervision. (a1) shows the ground truth 3D CT, (a2-a3) shows the transformed 3D CT by TransNet-v1 and TransNet-v2, respectively. (b1-b3) show the zoomed-in images of the two ROIs in red and green rectangles in (a1) for ground-truth, TransNet-v1, TransNet-v2, respectively. (c1-c3) shows one channel feature map extracted by a fixed pyramid level of SegNet from ground truth 3D CT,  the transformed 3D CT obtained via TransNet-v1, and the transformed 3D CT obtained via TransNet-v2, respectively. (d1-d3) show the corresponding zoomed-in ROIs. (e1) shows the input 2D projection. The angle for this projection is 60°. (e2) shows the line profiles of the yellow line (as shown in (a1)). The window level for all CT images are set to [-800, 800] HU.
	\end{figure}
	The quantitative evaluation results of the efficacy of perceptual supervision are listed in Table 1. It can be seen that with perceptual supervision, MAE was reduced from 136.0±21.0 HU to 108.1±9.1 HU averagely, NMAE was reduced from 0.043±0.008 to 0.035±0.006. In addition, PSNR was improved from 20.8±1.9 dB to 22.5±2.0 dB, and SSIM from 0.905±0.026 to 0.937±0.011. P-values were also calculated between TransNet-v1 and TransNet-v2 among 10 phases on these metrics for each patient, and the values were all smaller than 0.001, which means that TransNet-v2 significantly outperforms TransNet-v1 in deriving more accurate 3D volumetric images from a 2D project. 
	
	\begin{table}[htbp]
		\centering
		\caption{Numerical comparison result between TransNet-v1 and TransNet-v2 at projection angle 60°}
		\hspace*{-1cm}
		\begin{tabular}{ccccccccc}
			\hline
			\multirow{2}{*}{\begin{tabular}[c]{@{}c@{}}Patient \\ ID\end{tabular}} & \multicolumn{2}{c}{MAE (HU)} & \multicolumn{2}{c}{NMAE}  & \multicolumn{2}{c}{PSNR (dB)} & \multicolumn{2}{c}{SSIM}  \\ \cline{2-9} 
			& v1            & v2           & v1          & v2          & v1            & v2            & v1          & v2          \\ \hline
			1                                                                        & 117.4±2.9     & 106.9±2.0    & 0.044±0.002 & 0.04±0.002  & 20.0±0.6      & 20.9±0.6      & 0.925±0.003 & 0.939±0.002 \\
			2                                                                        & 122.9±1.6     & 106.8±0.9    & 0.042±0.001 & 0.036±0.001 & 20.3±0.3      & 21.8±0.4      & 0.915±0.002 & 0.940±0.001 \\
			3                                                                        & 130.2±2.2     & 103.7±0.5    & 0.038±0.000 & 0.03±0.001  & 22.3±0.2      & 24.2±0.3      & 0.917±0.001 & 0.945±0.001 \\
			4                                                                        & 124.0±8.5     & 98.6±3.0     & 0.04±0.002  & 0.032±0.002 & 20.9±0.9      & 22.6±1.0      & 0.913±0.005 & 0.941±0.002 \\
			5                                                                        & 138.8±4.4     & 103.6±2.4    & 0.034±0.001 & 0.025±0.001 & 24.0±0.2      & 25.9±0.1      & 0.921±0.005 & 0.946±0.002 \\
			6                                                                        & 134.1±0.6     & 117.4±0.2    & 0.051±0.001 & 0.045±0.001 & 18.5±0.1      & 19.5±0.1      & 0.911±0.001 & 0.931±0.000 \\
			7                                                                        & 132.6±1.0     & 108.5±0.7    & 0.053±0.001 & 0.044±0.001 & 18.4±0.2      & 20.0±0.2      & 0.913±0.001 & 0.942±0.001 \\
			8                                                                        & 179.4±0.5     & 122.4±0.3    & 0.044±0.000 & 0.030±0.000 & 22.2±0.0      & 25.0±0.0      & 0.869±0.001 & 0.926±0.000 \\
			9                                                                        & 113.2±1.6     & 93.9±0.5     & 0.035±0.001 & 0.029±0.001 & 21.7±0.5      & 23.2±0.5      & 0.929±0.001 & 0.949±0.000 \\
			10                                                                       & 144.2±4.6     & 116.1±1.9    & 0.045±0.001 & 0.036±0.002 & 20.1±0.4      & 22.0±0.5      & 0.882±0.005 & 0.924±0.003 \\
			11                                                                       & 132.2±4.0     & 108.3±0.5    & 0.039±0.002 & 0.032±0.003 & 21.8±0.7      & 23.5±0.9      & 0.911±0.003 & 0.938±0.001 \\
			12                                                                       & 125.7±3.2     & 105.7±1.8    & 0.048±0.002 & 0.040±0.001 & 19.0±0.3      & 20.5±0.3      & 0.917±0.003 & 0.943±0.002 \\
			13                                                                       & 145.0±2.0     & 109.1±0.9    & 0.037±0.001 & 0.027±0.000 & 23.2±0.2      & 25.2±0.1      & 0.908±0.003 & 0.938±0.001 \\
			14                                                                       & 162.6±1.1     & 117.9±0.5    & 0.040±0.000 & 0.029±0.000 & 22.2±0.0      & 24.6±0.0      & 0.883±0.001 & 0.929±0.001 \\
			15                                                                       & 118.3±2.6     & 98.8±1.5     & 0.043±0.002 & 0.036±0.002 & 19.8±0.5      & 21.4±0.5      & 0.919±0.003 & 0.945±0.001 \\
			16                                                                       & 171.2±1.4     & 122.6±0.7    & 0.042±0.000 & 0.030±0.000 & 22.1±0.1      & 24.5±0.0      & 0.878±0.002 & 0.926±0.001 \\
			17                                                                       & 143.3±1.3     & 113.1±0.4    & 0.048±0.001 & 0.038±0.001 & 20.1±0.2      & 22.1±0.2      & 0.899±0.001 & 0.936±0.000 \\
			18                                                                       & 106.8±6.0     & 93.4±2.1     & 0.035±0.004 & 0.031±0.004 & 21.9±1.3      & 23.0±1.5      & 0.936±0.003 & 0.949±0.002 \\
			19                                                                       & 177.8±2.6     & 125.6±1.0    & 0.068±0.001 & 0.048±0.000 & 16.2±0.1      & 18.7±0.2      & 0.826±0.003 & 0.903±0.001 \\
			20                                                                       & 116.5±1.0     & 102.4±0.7    & 0.046±0.001 & 0.040±0.000 & 19.6±0.1      & 20.5±0.1      & 0.928±0.001 & 0.943±0.001 \\ \hline
			Average                                                                  & 136.0±21.0    & 108.1±9.1    & 0.043±0.008 & 0.035±0.006 & 20.8±1.9      & 22.5±2.0      & 0.905±0.026 & 0.937±0.011 \\ \hline
		\end{tabular}
	\end{table}%
	 
	\noindent 
	\subsection{Efficacy of adversarial supervision}
	
	To demonstrate the efficacy of the adversarial supervision, we compared the transformation accuracy between TransNet-v2, which is supervised by a combination of MAE, GD and perceptual losses, and TransNet-v3, which is supervised by a combination of MAE, GD, perceptual and adversarial losses. This comparison was conducted on all the patients at projection angle = 90° (left-right projection view) for demonstration purpose.
	
	\begin{figure}
		\begin{center}		
		\noindent \includegraphics*[width=6.50in, height=4.20in, keepaspectratio=true]{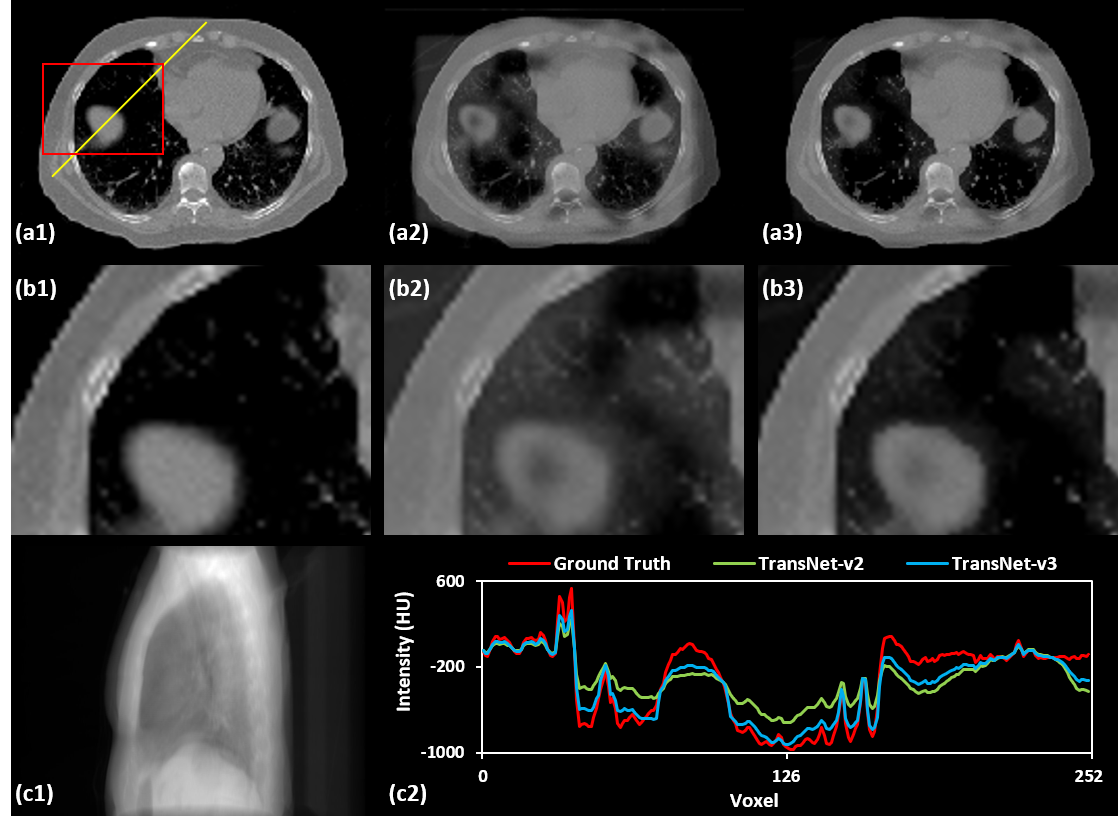}
		\end{center}
		
		\noindent Fig. 3. Visual comparison between TransNet-v2 and TransNet-v3 to demonstrate the efficacy of using adversarial supervision. (a1) shows the ground truth 3D CT. (a2) shows the transformed 3D CT by using TransNet-v2. (a3) shows the transformed 3D CT by using TransNet-v3. (b1-b3) show the zoomed-in ROI denoted by the red rectangle in a(1). (c1) shows the input 2D projection. The angle for this projection is 90°. (e2) shows the line profiles of the yellow line on the ground truth 3D CT and transformed 3D CT via two comparing methods. The window level for all CT images are set to [-800, 800] HU.
	\end{figure}

	Fig. 3 shows the visual comparison between TransNet-v2 and TransNet-v3 for a patient. An axial slice of the ground truth 3D CT, and the 3D CT generated by TransNet-v2 and TransNet-v3 are shown in subfigure (a1) - (a3), respectively, and the corresponding zoomed-in images are shown in (b1) - (b3). As can be seen from these images, TranNet-v2 seems to induce noise on the generated 3D images, particularly around the chest wall-lung interfaces. By using adversarial supervision, TransNet-v3 can effectively remove these artificial noises imposed by TranNet-v2. The performance of TransNet-v3 versus TransNet-v2 can be also seen from the line profile in (c2). To quantitatively evaluate the efficacy of the adversarial supervision, MAE, NMAE, PSNR and SSIM were calculated for the transformed 3D CT generated by TransNet-v2 and TransNet-v3, respectively, and shown in Table 2. Averaged over the 20 patients, MAE, NMAE, PSNR and SSIM of the 3DCT generated by TransNet-v2 are 110.1±15.2 HU, 0.035±0.008, 22.8±2.3 dB and 0.939±0.014, respectively. These metrics were improved by TransNet-v3 to 103.9±11.6 HU, 0.033±0.006, 23.0±1.9 and 0.944±0.009, respectively. P-values calculated on these metrics between TransNet-v2 and TransNet-v3 are all smaller than 0.001. These values have demonstrated that TransNet-v3 significantly outperforms TransNet-v2 in generating more realistic 3D CT images on most of the testing cases.
	
	\begin{table}[htbp]
		\centering
		\caption{Numerical comparison between transNet-v2 and transNet-v3 (projection angle = 90°).}
		\hspace*{-1cm}
		\begin{tabular}{ccccccccc}
			\hline
			\multirow{2}{*}{\begin{tabular}[c]{@{}c@{}}Patient \\ ID\end{tabular}} & \multicolumn{2}{c}{MAE (HU)} & \multicolumn{2}{c}{NMAE}  & \multicolumn{2}{c}{PSNR (dB)} & \multicolumn{2}{c}{SSIM}  \\ \cline{2-9} 
			& v2            & v3           & v2          & v3          & v2            & v3            & v2          & v3          \\ \hline
			1                                                                      & 106.5±3.6     & 101.4±2.9    & 0.04±0.003  & 0.038±0.002 & 21.1±0.7      & 21.5±0.7      & 0.941±0.003 & 0.946±0.003 \\
			2                                                                      & 103.3±1.1     & 99.2±0.7     & 0.035±0.001 & 0.034±0.001 & 22.2±0.4      & 22.4±0.3      & 0.944±0.001 & 0.947±0.001 \\
			3                                                                      & 102.8±0.6     & 113.7±0.4    & 0.030±0.001 & 0.033±0.001 & 24.4±0.3      & 23.9±0.3      & 0.945±0.001 & 0.938±0.001 \\
			4                                                                      & 105.9±1.7     & 93.9±3.5     & 0.034±0.003 & 0.03±0.002  & 22.8±1.0      & 23.3±1.0      & 0.942±0.001 & 0.949±0.002 \\
			5                                                                      & 88.3±0.5      & 104.5±0.5    & 0.022±0.000 & 0.026±0.000 & 27.3±0.1      & 26.0±0.0      & 0.961±0.001 & 0.946±0.000 \\
			6                                                                      & 128.9±0.3     & 114.6±0.5    & 0.049±0.000 & 0.044±0.001 & 19.4±0.1      & 20.0±0.1      & 0.928±0.001 & 0.938±0.000 \\
			7                                                                      & 116±0.3       & 110.1±0.4    & 0.047±0.001 & 0.044±0.001 & 19.5±0.2      & 20.1±0.2      & 0.930±0.0   & 0.941±0.000 \\
			8                                                                      & 135.3±0.8     & 126.5±0.6    & 0.033±0.000 & 0.031±0.000 & 24.5±0.0      & 25.1±0.0      & 0.915±0.001 & 0.930±0.001 \\
			9                                                                      & 102.8±1.7     & 86.9±1.1     & 0.032±0.002 & 0.027±0.001 & 23.1±0.5      & 23.9±0.5      & 0.947±0.001 & 0.957±0.001 \\
			10                                                                     & 138.2±3.0     & 107.3±1.4    & 0.043±0.002 & 0.033±0.001 & 21.4±0.5      & 22.7±0.4      & 0.91±0.004  & 0.936±0.002 \\
			11                                                                     & 97.1±1.2      & 100.1±1.8    & 0.029±0.003 & 0.03±0.003  & 24.5±0.9      & 24.0±0.9      & 0.951±0.000 & 0.945±0.001 \\
			12                                                                     & 102.8±0.8     & 104.2±1.0    & 0.039±0.001 & 0.039±0.001 & 21.3±0.2      & 21.1±0.2      & 0.951±0.001 & 0.95±0.001  \\
			13                                                                     & 99.3±1.5      & 102.3±1.1    & 0.025±0.000 & 0.026±0.000 & 26.1±0.1      & 25.8±0.1      & 0.949±0.001 & 0.946±0.001 \\
			14                                                                     & 118.4±0.7     & 109.3±0.5    & 0.029±0.000 & 0.027±0.000 & 24.9±0.1      & 25.4±0.0      & 0.933±0.001 & 0.941±0.000 \\
			15                                                                     & 94.2±1.3      & 86.8±1.6     & 0.034±0.002 & 0.031±0.002 & 22.1±0.4      & 22.3±0.5      & 0.953±0.001 & 0.956±0.001 \\
			16                                                                     & 112.3±0.8     & 109±0.4      & 0.028±0.000 & 0.027±0.000 & 25.4±0.1      & 25.4±0.0      & 0.941±0.001 & 0.942±0.000 \\
			17                                                                     & 119.2±0.3     & 119.2±0.3    & 0.040±0.001 & 0.04±0.001  & 21.8±0.3      & 21.8±0.2      & 0.927±0.000 & 0.928±0.001 \\
			18                                                                     & 95.5±4.1      & 82.6±4.5     & 0.031±0.005 & 0.027±0.005 & 23.2±1.6      & 24±1.7        & 0.951±0.002 & 0.960±0.003 \\
			19                                                                     & 142.1±0.7     & 120±1.0      & 0.054±0.001 & 0.046±0.001 & 19.1±0.2      & 19.8±0.2      & 0.914±0.001 & 0.926±0.001 \\
			20                                                                     & 105.7±0.8     & 95.2±0.8     & 0.042±0.000 & 0.037±0.000 & 20.5±0.1      & 21.4±0.1      & 0.941±0.001 & 0.953±0.001 \\ \hline
			Average                                                                & 110.1±15.2    & 103.9±11.6   & 0.035±0.008 & 0.033±0.006 & 22.8±2.3      & 23±1.9        & 0.939±0.014 & 0.944±0.009 \\ \hline
		\end{tabular}
	\end{table}%
	
	\subsection{Results at four different projection angles}
	
	Fig. 4 displays the 3D CT images derived by TransNet-v3 from a 2D projection data at four different projection angles (i.e., 0°, 30°, 60°, 90°). All of them show a good agreement with the ground truth images. The calculated MAE, NMAE, PSNR and SSIM are listed in Table 3-6, respectively. These numerical results have well demonstrated the robustness of our proposed method at different projection angles. It can be also observed that among these four angles, 0° projection angle (i.e., anterior-posterior projection view) yielded the best performance, while 30° projection angle (i.e., left anterior-oblique projection view) yielded the relatively worst performance.

	\begin{figure}
		\begin{center}
		\noindent \includegraphics*[width=6.50in, height=4.20in, keepaspectratio=true]{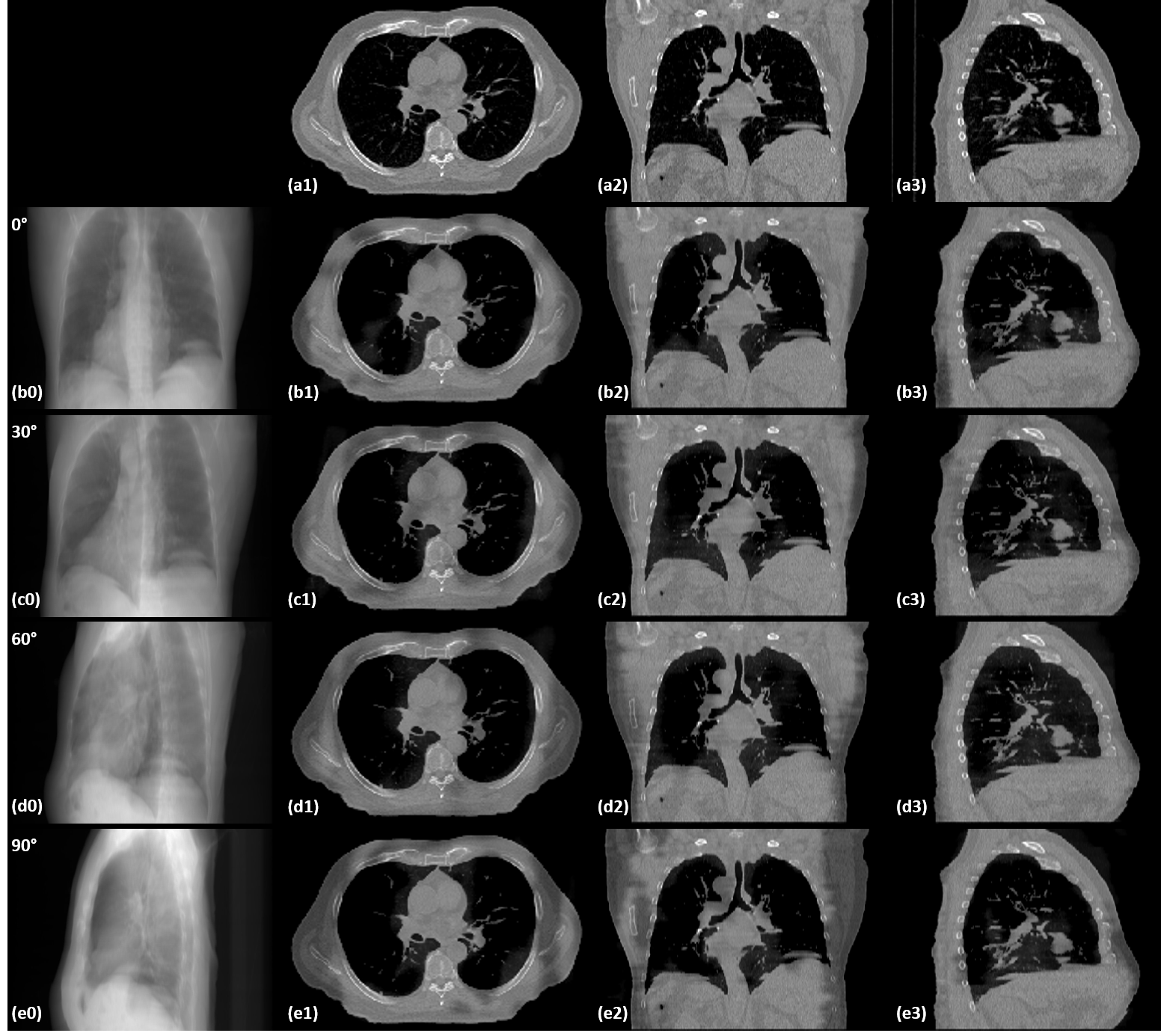}
		\end{center}
	
		\noindent Fig. 4. 3D CT images generated for a patient case by Trans-Net3 using a 2D projection at four different projection angles (i.e., 0°, 30°, 60°, 90°), respectively. (a1-a3) show the ground truth 3D CT in axial, coronal and sagittal views. (b0) shows the 2D projection at 0° projection angle. (b1-b3) show the three views of the 3D CT derived from this 2D projection by TransNet-v3. Row (c-d) show the 2D projection data and corresponding generated 3D CT images for projection angle 30°, 60°, 90°, respectively. The window level for all CT images are [-800, 800] HU.
	\end{figure}

	\begin{table}[htbp]
		\centering
		\caption{MAE (HU) results of TransNet-v3 for 4 different projection angles.}
		\begin{tabular}{cccccc}
			\hline
			\begin{tabular}[c]{@{}c@{}}Patient \\ ID\end{tabular}            & 0°        & 30°        & 60°       & 90°        & Average over angles \\ \hline
			1                                                                & 93.8±2.4  & 102.4±3.3  & 96.3±2.1  & 101.4±2.9  & 98.5±4.5            \\
			2                                                                & 78.9±1.4  & 99.2±1     & 99.6±0.9  & 99.2±0.7   & 94.2±9.0            \\
			3                                                                & 106.2±2.6 & 98.5±0.6   & 92.2±0.6  & 113.7±0.4  & 102.6±8.3           \\
			4                                                                & 71.3±3.0  & 101.3±1.6  & 87.9±3.4  & 93.9±3.5   & 88.6±11.6           \\
			5                                                                & 75.4±0.6  & 84.9±0.5   & 92.2±3.0  & 104.5±0.5  & 89.2±10.9           \\
			6                                                                & 99.7±0.3  & 123.3±0.3  & 107.2±0.3 & 114.6±0.5  & 111.2±9.0           \\
			7                                                                & 110.6±0.8 & 110.7±0.3  & 98.4±0.7  & 110.1±0.4  & 107.4±5.3           \\
			8                                                                & 114.9±0.9 & 128.3±0.7  & 118.7±0.6 & 126.5±0.6  & 122.1±5.6           \\
			9                                                                & 72.1±0.8  & 98.3±1.6   & 84.5±0.4  & 86.9±1.1   & 85.5±9.5            \\
			10                                                               & 106.5±1.6 & 131.9±2.8  & 104.4±1.7 & 107.3±1.4  & 112.5±11.5          \\
			11                                                               & 85.5±1.6  & 93.4±1.2   & 95.6±0.7  & 100.1±1.8  & 93.6±5.5            \\
			12                                                               & 100.1±0.8 & 98.4±0.8   & 93.9±1.8  & 104.2±1.0  & 99.1±3.9            \\
			13                                                               & 84±1.2    & 95.4±1.4   & 96.9±0.6  & 102.3±1.1  & 94.6±6.8            \\
			14                                                               & 89.6±0.3  & 113.2±0.7  & 104.4±0.4 & 109.3±0.5  & 104.1±9.1           \\
			15                                                               & 65.9±1.6  & 90.3±1.2   & 90.7±1.2  & 86.8±1.6   & 83.4±10.4           \\
			16                                                               & 81.9±0.7  & 107.4±0.7  & 111.2±0.5 & 109±0.4    & 102.3±12.2          \\
			17                                                               & 115.8±0.9 & 113.6±0.3  & 100.7±0.3 & 119.2±0.3  & 112.3±7.1           \\
			18                                                               & 71.8±3.1  & 91.8±3.8   & 82.7±2.3  & 82.6±4.5   & 82.2±7.9            \\
			19                                                               & 96.4±0.7  & 135±0.6    & 112.7±0.9 & 120±1.0    & 116±14.1            \\
			20                                                               & 92.4±0.7  & 101.3±0.8  & 92.6±0.6  & 95.2±0.8   & 95.3±3.7            \\ \hline
			\begin{tabular}[c]{@{}c@{}}Average over \\ patients\end{tabular} & 90.6±15.5 & 105.3±14.1 & 97.5±9.2  & 103.9±11.6 & 99.3±14.1           \\ \hline
		\end{tabular}
	\end{table}%
	
	\begin{table}[htbp]
		\centering
		\caption{NMAE results of TransNet-v3 for 4 different projection angles.}
		\begin{tabular}{cccccc}
			\hline
			\begin{tabular}[c]{@{}c@{}}Patient\\ ID\end{tabular}             & 0°          & 30°         & 60°         & 90°         & Average over angles \\ \hline
			1                                                                & 0.035±0.002 & 0.038±0.003 & 0.036±0.002 & 0.038±0.002 & 0.037±0.003         \\
			2                                                                & 0.027±0.001 & 0.034±0.001 & 0.034±0.001 & 0.034±0.001 & 0.032±0.003         \\
			3                                                                & 0.031±0.001 & 0.029±0.001 & 0.027±0.001 & 0.033±0.001 & 0.030±0.003         \\
			4                                                                & 0.023±0.002 & 0.032±0.003 & 0.028±0.002 & 0.030±0.002 & 0.028±0.004         \\
			5                                                                & 0.019±0.000 & 0.021±0.000 & 0.023±0.001 & 0.026±0.00  & 0.022±0.003         \\
			6                                                                & 0.038±0.001 & 0.047±0.000 & 0.041±0.000 & 0.044±0.001 & 0.043±0.003         \\
			7                                                                & 0.044±0.001 & 0.044±0.001 & 0.04±0.001  & 0.044±0.001 & 0.043±0.002         \\
			8                                                                & 0.028±0.000 & 0.032±0.000 & 0.029±0.000 & 0.031±0.00  & 0.030±0.001         \\
			9                                                                & 0.023±0.001 & 0.031±0.002 & 0.026±0.001 & 0.027±0.001 & 0.027±0.003         \\
			10                                                               & 0.033±0.001 & 0.041±0.002 & 0.032±0.001 & 0.033±0.001 & 0.035±0.004         \\
			11                                                               & 0.025±0.002 & 0.028±0.003 & 0.028±0.002 & 0.03±0.003  & 0.028±0.003         \\
			12                                                               & 0.038±0.001 & 0.037±0.001 & 0.035±0.001 & 0.039±0.001 & 0.037±0.002         \\
			13                                                               & 0.021±0.000 & 0.024±0.000 & 0.024±0.000 & 0.026±0.000 & 0.024±0.002         \\
			14                                                               & 0.022±0.000 & 0.028±0.000 & 0.026±0.000 & 0.027±0.000 & 0.026±0.002         \\
			15                                                               & 0.024±0.001 & 0.032±0.001 & 0.033±0.002 & 0.031±0.002 & 0.030±0.004         \\
			16                                                               & 0.02±0.000  & 0.026±0.000 & 0.027±0.000 & 0.027±0.000 & 0.025±0.003         \\
			17                                                               & 0.039±0.001 & 0.038±0.001 & 0.034±0.001 & 0.04±0.001  & 0.038±0.003         \\
			18                                                               & 0.024±0.004 & 0.03±0.005  & 0.027±0.004 & 0.027±0.005 & 0.027±0.005         \\
			19                                                               & 0.037±0.001 & 0.051±0.001 & 0.043±0.000 & 0.046±0.001 & 0.044±0.005         \\
			20                                                               & 0.036±0.000 & 0.04±0.000  & 0.036±0.000 & 0.037±0.000 & 0.038±0.001         \\ \hline
			\begin{tabular}[c]{@{}c@{}}Average over \\ patients\end{tabular} & 0.029±0.008 & 0.034±0.008 & 0.031±0.006 & 0.033±0.006 & 0.032±0.007         \\ \hline
		\end{tabular}
	\end{table}%

	\begin{table}[htbp]
		\centering
		\caption{PSNR (dB) results of TransNet-v3 for 4 different projection angles.}
		\begin{tabular}{cccccc}
			\hline
			\begin{tabular}[c]{@{}c@{}}Patient\\ ID\end{tabular}             & 0°       & 30°      & 60°      & 90°      & Average over angles \\ \hline
			1                                                                & 22.0±0.7 & 21.4±0.7 & 21.8±0.6 & 21.5±0.7 & 21.7±0.7            \\
			2                                                                & 23.9±0.4 & 22.5±0.4 & 22.3±0.4 & 22.4±0.3 & 22.8±0.7            \\
			3                                                                & 24.3±0.3 & 24.8±0.3 & 25.1±0.3 & 23.9±0.3 & 24.5±0.6            \\
			4                                                                & 25.1±1.0 & 23.2±1.0 & 23.9±1.0 & 23.3±1   & 23.9±1.3            \\
			5                                                                & 28.4±0.1 & 27.6±0.1 & 26.9±0.3 & 26±0.0   & 27.2±0.9            \\
			6                                                                & 20.9±0.1 & 19.8±0.1 & 20.5±0.1 & 20±0.1   & 20.3±0.5            \\
			7                                                                & 20.1±0.2 & 19.9±0.2 & 21.0±0.2 & 20.1±0.2 & 20.3±0.5            \\
			8                                                                & 25.8±0.1 & 24.9±0.0 & 25.6±0.0 & 25.1±0.0 & 25.4±0.4            \\
			9                                                                & 25.1±0.5 & 23.4±0.5 & 24.2±0.5 & 23.9±0.5 & 24.2±0.8            \\
			10                                                               & 23.0±0.4 & 21.8±0.5 & 23.0±0.4 & 22.7±0.4 & 22.6±0.7            \\
			11                                                               & 25.2±0.9 & 24.8±0.9 & 24.6±0.9 & 24±0.9   & 24.7±1.0            \\
			12                                                               & 21.6±0.2 & 21.6±0.2 & 21.8±0.3 & 21.1±0.2 & 21.5±0.3            \\
			13                                                               & 27.1±0.1 & 26.4±0.1 & 26.3±0.1 & 25.8±0.1 & 26.4±0.5            \\
			14                                                               & 26.6±0.0 & 25.3±0.1 & 25.8±0.0 & 25.4±0.0 & 25.8±0.5            \\
			15                                                               & 24.3±0.5 & 22.5±0.4 & 22.2±0.5 & 22.3±0.5 & 22.8±1.0            \\
			16                                                               & 27.6±0.1 & 25.7±0.1 & 25.4±0.0 & 25.4±0.0 & 26.1±0.9            \\
			17                                                               & 22.3±0.2 & 22.3±0.3 & 23.1±0.2 & 21.8±0.2 & 22.4±0.5            \\
			18                                                               & 25.0±1.6 & 23.5±1.6 & 24.1±1.5 & 24.0±1.7 & 24.2±1.6            \\
			19                                                               & 21.2±0.2 & 19.5±0.2 & 20.1±0.2 & 19.8±0.2 & 20.2±0.7            \\
			20                                                               & 21.4±0.1 & 20.8±0.1 & 21.5±0.1 & 21.4±0.1 & 21.3±0.3            \\ \hline
			\begin{tabular}[c]{@{}c@{}}Average over \\ patients\end{tabular} & 24.1±2.4 & 23.1±2.2 & 23.5±2.0 & 23.0±1.9 & 23.4±2.2            \\ \hline
		\end{tabular}
	\end{table}%

	\begin{table}[htbp]
		\centering
		\caption{SSIM results of TransNet-v3 for 4 different projection angles.}
		\begin{tabular}{cccccc}
			\hline
			\begin{tabular}[c]{@{}c@{}}Patient\\ ID\end{tabular}             & 0°          & 30°         & 60°         & 90°         & Average over angles \\ \hline
			1                                                                & 0.953±0.002 & 0.945±0.003 & 0.95±0.002  & 0.946±0.003 & 0.949±0.004         \\
			2                                                                & 0.964±0.001 & 0.948±0.001 & 0.946±0.001 & 0.947±0.001 & 0.951±0.007         \\
			3                                                                & 0.942±0.002 & 0.949±0.001 & 0.955±0.001 & 0.938±0.001 & 0.946±0.007         \\
			4                                                                & 0.968±0.001 & 0.947±0.001 & 0.956±0.002 & 0.949±0.002 & 0.955±0.008         \\
			5                                                                & 0.907±0.001 & 0.964±0.001 & 0.957±0.003 & 0.946±0.000 & 0.959±0.009         \\
			6                                                                & 0.951±0.000 & 0.934±0.001 & 0.945±0.000 & 0.938±0.000 & 0.942±0.007         \\
			7                                                                & 0.940±0.000 & 0.937±0.000 & 0.953±0.001 & 0.941±0.000 & 0.943±0.006         \\
			8                                                                & 0.942±0.001 & 0.923±0.001 & 0.938±0.001 & 0.930±0.001 & 0.933±0.007         \\
			9                                                                & 0.968±0.000 & 0.951±0.001 & 0.960±0.000 & 0.957±0.001 & 0.959±0.006         \\
			10                                                               & 0.940±0.001 & 0.918±0.003 & 0.941±0.002 & 0.936±0.002 & 0.934±0.010         \\
			11                                                               & 0.959±0.001 & 0.954±0.000 & 0.952±0.001 & 0.945±0.001 & 0.953±0.005         \\
			12                                                               & 0.955±0.001 & 0.955±0.000 & 0.958±0.001 & 0.950±0.001 & 0.954±0.003         \\
			13                                                               & 0.961±0.001 & 0.953±0.001 & 0.952±0.000 & 0.946±0.001 & 0.953±0.005         \\
			14                                                               & 0.958±0.000 & 0.939±0.001 & 0.947±0.000 & 0.941±0.000 & 0.946±0.007         \\
			15                                                               & 0.973±0.001 & 0.957±0.001 & 0.955±0.001 & 0.956±0.001 & 0.960±0.008         \\
			16                                                               & 0.967±0.001 & 0.945±0.001 & 0.942±0.000 & 0.942±0.000 & 0.949±0.011         \\
			17                                                               & 0.934±0.001 & 0.934±0.000 & 0.949±0.000 & 0.928±0.001 & 0.936±0.008         \\
			18                                                               & 0.969±0.002 & 0.954±0.002 & 0.961±0.001 & 0.960±0.003 & 0.961±0.006         \\
			19                                                               & 0.949±0.001 & 0.922±0.001 & 0.932±0.001 & 0.926±0.001 & 0.932±0.010         \\
			20                                                               & 0.953±0.001 & 0.946±0.001 & 0.955±0.001 & 0.953±0.001 & 0.952±0.004         \\ \hline
			\begin{tabular}[c]{@{}c@{}}Average over \\ patients\end{tabular} & 0.956±0.012 & 0.944±0.012 & 0.951±0.008 & 0.944±0.009 & 0.949±0.012         \\ \hline
		\end{tabular}
	\end{table}%

	\bigbreak
	
	\noindent 
	\section{Discussion}
	
	Shen et al. proposed to employ a deep neural network, for the first time, to learn the mapping relationship between a single 2D projection radiograph of a patient to the corresponding 3D anatomy, and used this trained network to generate volumetric tomographic X-ray images of the patient from a single anterior-posterior projection view \cite{RN1}. In their experiments, they tested on one patient’s lung 4D CT images and achieved 0.025 NMAE, 27.157 dB PSNR, and 0.838 SSIM. Inspired by this work, we have employed a more advanced GAN network for this ill-posted 2D-to-3D transformation problem, and integrated perceptual supervision into it. As demonstrated in Fig.2 and Table 1, adding the perceptual supervision to our network can significantly enhance its sensitivity for accurate lung boundary regression. In addition, the adversarial supervision significantly improved the realism of the generated 3D images in relative to the ground truth images, as demonstrated in Fig.3 and Table 2. The efficacy and consistency of our method at different projection angles have also been well demonstrated by our experimental results in Fig. 4 and Table 3-6. At the anterior-posterior projection angle, the NMAE value achieved by our method for the 20 patient cases ranges from 0.019 to 0.044, with a mean value of 0.029 and a standard deviation of 0.008. The achieved PSNR value ranges from 20.1 to 28.4 dB, with a mean value of 24.1 dB and a standard deviation of 2.4 dB. The achieved SSIM value ranges from 0.907 to 0.973, with a mean value of 0.956 and a standard deviation of 0.012. Although it is difficult to fairly compare our method with Shen’s method since 20 different patient cases were used in our study, we can observe that our method achieves comparable performance for the anterior-posterior projection angle. In our study, we have also found that the anterior-posterior projection angle (i.e., 0°) yielded the best performance among the four testing projection angles. This might be due to the thinnest patient thickness at this angle, which results in more pixels of useful information with less superimposition on each pixel and hence leads to a less ill-posed problem. 
	
	In this preliminary study, we pre-trained the SegNet segmentation network only for lung to see whether the perceptual supervision can help improve the accuracy of lung boundary on the generated 3D images. With the obvious efficacy of this perceptual supervision being observed from our experimental results, in future we will extend our perceptual supervision to other moving organs, such as heart and esophagus, and tumor for more accurate image boundary of these moving objects. 
	
	This proof-of-concept study has demonstrated the feasibility of deriving a 3D volumetric image from a single 2D projection of different projection angle, paving the way for in-treatment real-time volumetric imaging using the on-board imager. However, there are still two challenges that need to be overcome. First, without any projection images and corresponding 3D images available before the actual treatment, we have to use the patient’s 4D planning CT images (i.e., 3D CT images at each phase) and simulated projection data to train the patient-specific networks. These simulated projections for training need to have similar image quality compared to the actual in-treatment 2D kV projections that will be used to drive the instantaneous 3D images. Conventional ray-tracing methods usually yield ideal projections, while actual kV projections acquired by flat panel detector are contaminated by scatters and noises. This disparity can be addressed by Monte Carlo simulation, which can simulate all possible physics events for each imaging photon to yield realistic projection data for training. Another challenge is the limited size of the on-board imager, which may not cover the entire contralateral lung for patients of relatively large body size, leading to a limited field of view (FOV) in the generated 3D image. This limited FOV might be extended using the 4D CT images and a lung motion model as a prior knowledge. We will focus our future work on the investigation of the potential solutions to address these two issues.

	\bigbreak
	
	\noindent 
	\section{Conclusion}
	
	In summary, we have proposed a novel perceptual supervision-based GAN method to derive instantaneous volumetric image from a single projection. The efficacy and robustness of our method have been tested for different projection angles in this study. By validating the feasibility of deriving more informative 3D image from a single 2D projection image at different projection angle via our proposed method, this proof-of-concept study paves the way for in-treatment on-board real-time volumetric imaging for more accurate and active motion management during lung SBRT treatment.

	\noindent 
	\bigbreak
	{\bf ACKNOWLEDGEMENT}
	
	This research is supported in part by the National Cancer Institute of the National Institutes of Health under Award Number R01CA215718, and Dunwoody Golf Club Prostate Cancer Research Award, a philanthropic award provided by the Winship Cancer Institute of Emory University.

	\noindent 
	\bigbreak
	{\bf Disclosures}
	
	The authors declare no conflicts of interest.

	\noindent 
	
	\bibliographystyle{plainnat}  
	\bibliography{arxiv}      
	
\end{document}